\documentclass[a4paper,10pt]{article}
\usepackage{makeidx}
\usepackage[cp1251]{inputenc}
\usepackage{cite,amsmath,amsfonts,amsthm,fullpage}
\usepackage{youngtab}
\usepackage[usenames,dvipsnames]{color}
\usepackage{ulem}

\theoremstyle{plain}

\newtheorem{Lemma}{Lemma}
\newtheorem{Proposition}{Proposition}
\newtheorem{Corollary}{Corollary}
\newtheorem{Remark}{Remark}
\theoremstyle{remark}

\def\det{\mathrm {det}}

\def\bp{\begin{Proposition}}
\def\ep{\end{Proposition}}
\def\bc{\begin{Corollary}}
\def\ec{\end{Corollary}}
\def\bl{\begin{Lemma}}
\def\el{\end{Lemma}}
\def\be{\begin{equation}}
\def\ee{\end{equation}}
\def\br{\begin{Remark}\rm\small}
\def\er{\end{Remark}}
\def\brs{\begin{remarks}.\\ \rm\
\begin{enumerate}}
\def\ers{\end{enumerate}\end{remarks}}
\def\bea{\begin{eqnarray}}
\def\eea{\end{eqnarray}}

\def\det{\mathrm {det}}

\def\&{&{\hskip -20pt}}

\newcount\YDcount\YDcount=0
\def\YDsize{10pt}
\def\YD#1{\ifnum#1=0
 \ifnum\YDcount=0 \ifx\varnothing\undefined\emptyset\else\varnothing\fi
 \else\vskip1.4pt\egroup\YDcount=0\fi
\else
 \ifnum\YDcount=0 \YDcount=1\vcenter\bgroup\vskip1pt
 \else\nointerlineskip\fi
 \vbox{\hrule\hbox{\vrule height\YDsize
 \loop\hskip\YDsize\vrule\ifnum\YDcount<#1\advance\YDcount1\repeat}\hrule
 \kern-0.4pt}\expandafter\YD
\fi}

\begin{document}

\author{ E.N.Antonov \thanks{%
Petersburg Nuclear Physics Institute, Gatchina, RU-188350 St.Petersburg,
Russia, email: antonov@thd.pnpi.spb.ru} \and A. Yu. Orlov\thanks{%
Institute of Oceanology, Nahimovskii Prospekt 36, Moscow 117997, Russia, and
National Research University Higher School of Economics, International
Laboratory of Representation Theory and Mathematical Physics, 20
Myasnitskaya Ulitsa, Moscow 101000, Russia, email: orlovs@ocean.ru }}
\title{Instantons in $\sigma$ model and tau functions. A remark.}
\maketitle

\begin{abstract}
We show that a number of multiple integrals may be viewed as tau functions of
various integrable hierarchies. The instanton contributions in the
two-dimensional O(3)$\ \sigma $ model is an example of such an approach.
\end{abstract}

\numberwithin{equation}{section}

\section{Introduction}

The purpose of this paper is to interpretate the the contribution of
instantons in Euclidean Green function of the O(3) non-linear $\sigma $
model (or the continuum classical Heisenberg ferromagnet in two space
dimensions)\ in terms of tau functions of integrable hierarchies. This model
can be described by the action
\begin{equation}
S=\frac{1}{2f}\int \sum\limits_{a=1}^{3}\left( \partial _{\mu }\sigma
^{a}\left( x\right) \right) ^{2}  \label{1}
\end{equation}%
where $\sigma^{a},\, a=1,2,3$ are the components of the unit vector:
 $\sum\limits_{a=1}^{3}\sigma ^{a}\left( x\right) \sigma ^{a}\left(
x\right) =1$ ; $\mu =0,1.$

The model similar to a Yang-Mills theory and posesses exact multi-instanton
solutions. The Euclidean Green functions can be represented in the form
\begin{equation}
\frac{\int \phi \left( \sigma \right) \exp \left( -S\right)
\prod\limits_{x}d\sigma \left( x\right) }{\int \exp \left( -S\right)
\prod\limits_{x}d\sigma \left( x\right) }  \label{2}
\end{equation}%
Here $\phi \left( \sigma \right) $ is an arbitrary functional of $\sigma $.
If we parametrize $\sigma \left( x\right) $ with use of the complex function
\be\label{stereo}
\omega(z) =\frac{\sigma ^{1}(z)+i\sigma ^{2}(z)}{1+\sigma ^{3}(z)}
\ee
(the stereographic projection) obtained from the
field\ $\left( \sigma ^{1},\sigma ^{2},\sigma ^{3}\right) $\ and the complex
variable $z=x_{0}+ix^{1}$ instead of the time and space coordinate $x_{0%
\text{ }}$, $x_{1}$, then the instanton is the solution of the equation $%
\delta S=0$ with the topological charge $q>0$ is given \cite{BP}
\begin{equation}
\omega_q \left(a,b, z\right) =c\frac{\left( z-a_{1}\right) ...\left(
z-a_{q}\right) }{\left( z-b_{1}\right) ...\left( z-b_{q}\right) }  \label{3}
\end{equation}%
where $c$, $a_{i}$, $b_{i}$ are arbitrary complex parameters.

Let us note that the classical $\sigma$-model in Minkowski space 
is the well studied integrable model, see \cite{ZakharovNovikov-ed}.

\section{ The instanton contribution and the $\protect\tau $ function}

In \cite{FFS} the instanton contribution to  the the Eucledean Green functions of the fields $\sigma$ using the 
steepest descent aproximation was obtained. If $\phi$ is a functional of the instanton fields $\omega$, then
the evaluation of the functional integral around the instanton vacuums yields \cite{FFS} the answers
written in form of multiple integrals over instanton parameters
\begin{equation}
\left< \phi \right>_{\rm inst}=\left[\frac{\sum_{q\geq 0}\frac{K^{q}}{(q!)^{2}}\int \phi(\omega_q )
\prod_{i<j\leq q}\frac{ |a_{i}-a_{j}|^{2}|b_{i}-b_{j}|^{2}}{
|a_{i}-b_{j}|^{2}|b_{i}-a_{j}|^{2}}\prod_{i=1}^{q}  \frac{d^2 a_i d^2 b_i}{|a_i-b_i|^2}}   %
{\sum_{q\geq 0}\frac{K^{q}}{(q!)^{2}}\int 
\prod_{i<j\leq q}\frac{ |a_{i}-a_{j}|^{2}|b_{i}-b_{j}|^{2}}{
|a_{i}-b_{j}|^{2}|b_{i}-a_{j}|^{2}}\prod_{i=1}^{q}  \frac{d^2 a_i d^2 b_i}{|a_i-b_i|^2}}
\right]_{\rm reg},
\label{4}
\end{equation}%
where $K$ is a real constant obtained as the result of the regularization 
procedure\footnote{According to \cite{FFS} the constant $K$ is proportional to 
$k_0 f^{-2}_{\rm phys} 
\exp \left(-4\pi f^{-1}_{\rm phys}\right)\nu$ where $\nu$ is the substraction point,
$f_{\rm phys}$ is a physical coupling constant, $k_0$ is a constant depending on the cutoff method. }
and 
where for each $q$ the instanton solution $\omega $ is given by (\ref{3}).
The denominator in (\ref{4}) coincides with the partition
function $\Xi $\ of the neutral classical two-dimensional Coulomb system (CCS) in the grand
canonical ensemble with the definite temperature T (T=1 see \cite{FFS}) (such system was called 
the system of instanton quarks in \cite{FFS}). The point T=1 is above the critical temperature
(which is about T=1/2), this means that the Coulomb gas is in the plasma phase. (Below the critical
temperature the Coulomb particles form dipoles).
The symbol $[]_{\rm reg}$ means
that this expression should be regularized in the ultraviolet limit where $a_i \to b_j$.  Physical answers 
do not depend on the way of the regularization. 

Note that in fact, instanton-anti-instanton is also significant, see {LipatovBukhvostov} 
\cite{LipatovBukhvostov}, but this 
was not considered in the work
  \cite{FFS}, and we also will not touch on this much more involved topic.

\paragraph{Regularization.}
Let us notice that the answer (\ref{4}) was obtained \cite{FFS} as the result of the calculation of the functional 
integral and a certain regularization procedure and, in turn, the multiple integrals in (\ref{4}) are both 
infrared (IR) and ultroviolet (UV) divergent
and one needs an additional regularization procedure. 
In short it is discussed in \cite{FFS}, page 11. 

As for the IR divergency (the divergency in the limit $a_i,b_i\to \infty$)
 it just means that one
should be interested in the densities of the instanton partition function (and of the correlation function)
rather than the partition and the correlation functions themselves.
Then it is reasonable to restrict the domain of the integration over each $a_i$ to the $D=L\times L$ box in the 
complex plane, the same for $b_i$ \cite{FFS}. To get the density we divide each integral over $L^2$,
sumulteniousely we send the constant $K$ to $KL^2$.
Another way to get rid of IR divergency is to localise the problem is the additing the 

As for the ultraviolet regularization in the regions $b_i \approx a_j$ there are different ways:

(A) We can do the 
following: we produce the replacement $b_i \to b_i +\epsilon,\, {\bar b}_i \to {\bar b}_i -\epsilon$, 
where $\epsilon$ is a small real number where $\epsilon^{-1}$ may be treated as a cutoff in the momentum space.

In particular, for the one-instanton partition function we get 
\be\label{UVreg-A}
 K\int \frac{d^2a d^2 b}{|b-a|^2}\, \to \, \left(KL^2\right) L^{-2}\int_{D^2}
 \frac{d^2a d^2 b}{|a-b|^2_\epsilon }\,\quad{\rm where}\quad
 |a-b|^2_\epsilon : =|a-b|^2 -\epsilon^2+i\epsilon \Im (a-b)
\ee
The contribution of the region $b \approx a$ is finite and of order $\epsilon^{-1}$. Let us notice
that thanks to the structure of the numerators inside the integrals in (\ref{4}), the order of the $
q$-instanton integral is $\epsilon^{-q}$.

(B) One is to replace integrals by sums that is to consider the Coulomb gas on the 2D lattice
as mentioned in \cite{FFS} with the list of references. We can do it as follows: we take a small
real number $h$ (square grid spacing), and set 
\be\label{a_nmb_nm}
a(n,m)=nh+imh\,\quad b(n,m)=(n+\gamma )h+i(m+\gamma' )h
\ee
with non-integer $\gamma, \gamma'$.
In fact we have two lattices: one for positive, the other for negative Coulomb particles
\be\label{UVreg-B}
 K\int \frac{d^2a d^2 b}{|b-a|^2}\quad \to \quad \left(KL^2\right) L^{-2}\sum_{0\le n,n',m,m'\le L}
 \frac{h^{-2}}{|n'-n+im'-im+\frac 12(\gamma +i\gamma')|^2 }
\ee
The summation range $0\le n,m \le L$ will be also denoted $D$ as in the previous case.

Our goal is to relate (\ref{4}) with the regularizations (A)-(B) to classical integrable systems.

\section{Tau functions}

\subsection{Two-component KP and the regularization (A)\label{2KP-regA}}

In this case we use (\ref{UVreg-A}) and write the correlation function as
\begin{equation}
\left< \phi \right>_{\rm inst}^A=
\frac{\sum_{q\geq 0}\frac{K^{q}}{(q!)^{2}}\int_{D^{2q}} \phi(\omega_q )
\prod_{i<j\leq q}\frac{ |a_{i}-a_{j}|^{2}|b_{i}-b_{j}|^{2}}
{|a_{i}-b_{j}|^{2}_\epsilon |b_{i}-a_{j}|^{2}_\epsilon}
\prod_{i=1}^{q}  
\frac{d^2 a_i d^2 b_i}{|a_i-b_i|^2_\epsilon } }
{\sum_{q\geq 0}\frac{K^{q}}{(q!)^{2}}\int_{D^{2q}}
\prod_{i<j\leq q}\frac{ |a_{i}-a_{j}|^{2}|b_{i}-b_{j}|^{2}}
{|a_{i}-b_{j}|^{2}_\epsilon |b_{i}-a_{j}|^{2}_\epsilon}
\prod_{i=1}^{q}  
\frac{d^2 a_i d^2 b_i}{|a_i-b_i|^2_\epsilon }}%
\label{4A}
\end{equation}%

Compare with the $\tau $ function of the two-component KP (about
two-component KP see \cite{JM}) in terms of free fermion formalizm
\begin{equation}
\psi ^{(\alpha )}(z)=\sum_{i\in \mathbb{Z}}\psi _{i}^{(\alpha )}z^{i},\quad
\psi ^{\dagger (\alpha )}(z)=\sum_{i\in \mathbb{Z}}\psi _{i}^{\dagger
(\alpha )}z^{-1-i}  \label{5}
\end{equation}%
where $\alpha $ is a sort of fermions\ ($\alpha =1,2)$ and anti-commutators
companents $\psi _{i}^{(\alpha )},\psi _{i}^{\dagger (\alpha )}$\ are%
\begin{equation}
\left[ \psi _{i}^{(\alpha )},\psi _{j}^{(\beta )}\right] _{+}=0\qquad \left[
\psi _{i}^{\dagger (\alpha )},\psi _{j}^{\dagger (\beta )}\right] =0\qquad %
\left[ \psi _{i}^{(\alpha )},\psi _{j}^{\dagger (\beta )}\right] _{+}=\delta
_{\alpha ,\beta }\delta _{i,j}  \label{6}
\end{equation}%
The fermionic states with occupied level up to $n^{(1)},n^{(2)}$ \ satisfy the conditions%
\begin{equation*}
\langle n^{(1)},n^{(2)} | m^{(1)},m^{(2)} \rangle =\delta _{n^{(1)},m^{(1)} }\delta
_{n^{(2)},m^{(2)} }
\end{equation*}%
\begin{equation}
\psi _{i}^{(\alpha )}|n^{(\alpha )},*\rangle =\langle n^{(\alpha
)},*|\psi _{i}^{\dagger (\alpha )}= \psi _{-1-i}^{\dagger (\alpha
)}|n^{(\alpha )},*\rangle = \langle n^{(\alpha )},*|\psi _{-1-i}^{(\alpha
)}=0,\quad i<n^{(\alpha )}  \label{7}
\end{equation}%
The family of the $\tau $ functions of the two-component KP which is related to (\ref{4A}), is given by
\[
\tau (n,n^{(1)},n^{(2)},t^{(1)},t^{(2)})=
\]
\begin{equation}
\langle n^{(1)},n^{(2)}|\,\Gamma \left(
t^{(1)}\right)\Gamma\left(t^{(2)}\right) \,e^{K^{\frac{1}{2}}\int_{D^2} \psi ^{(1)}(a)\psi ^{\dag
(2)}({\bar{a}})d^2 a }\,e^{K^{\frac{1}{2}%
}\int_{D^2} \psi ^{(2)}({\bar{b}}-\epsilon)\psi ^{\dag (1)}(b+\epsilon)d^2 b%
 }\,|n^{(2)}-n^{(0)},n^{(1)}+n^{(0)}\rangle .  \label{8}
\end{equation}%
where the function
\begin{equation}
\Gamma \left( t^{(\alpha)}\right) =e^{\sum_{i>0} t_{i}^{(\alpha )}J_{i}^{(\alpha )}}  \label{9}
\end{equation}%
is expressed in terms of currents $J_{i}^{(\alpha )}$ by
\begin{equation}
:\psi ^{(\alpha )}(z)\psi ^{\dag (\alpha )}(z):=\psi ^{(\alpha )}(z)\psi
^{\dag (\alpha )}(z)-\langle 0|\psi ^{(\alpha )}(z)\psi ^{\dag (\alpha
)}(z)|0\rangle =\sum_{i\in \mathbb{Z}}\,J_{i}^{(\alpha )}\,z^{i-1}
\label{10}
\end{equation}%
or for the fermionic components we set for$\ m\in \mathbb{Z}$%
\begin{equation}
J_{m}^{(\alpha )}=\sum_{i\in \mathbb{Z}}:\psi _{i}^{(\alpha )}\psi
_{i+m}^{(\alpha )}:  \label{11}
\end{equation}%
Then we have the Heisenberg algebra commutation relation
\begin{equation}
\left[ J_{k}^{(\alpha )},J_{m}^{(\beta )}\right] =k\delta _{\alpha ,\beta
}\delta _{k+m,0}\quad .  \label{12}
\end{equation}%
The discrete variables $n^{(0)},n^{(\alpha)}$ and complex parameters $t^{(\alpha)}$, $\alpha=1,2$ are called 
the higher times of the two-component KP hierarchy. In what follows we put $n=0$.

Let us denote the sets $(n^{(\alpha)},t_1^{(\alpha)},t_2^{(\alpha)},\dots)$, $\alpha =1,2$ as ${\bf t}^{\alpha}$.

As the result of direct evaluation (using the relations in Appendix \ref{useful-relations}) we obtain the 
following $\tau $ function of the two-component KP:
\begin{equation}
\tau^A \left({\bf t}^{1},{\bf t}^{2}|D,\epsilon\right)=
\sum_{q\geq 0}\frac{K^{q}}{(q!)^{2}}\int_{D^{2q}}
\Phi _q\left(a,b,{\bf t}^{1},{\bf t}^{2}\right)
\prod_{i<j\leq q}\frac{ |a_{i}-a_{j}|^{2}|b_{i}-b_{j}|^{2}}
{|a_{i}-b_{j}|^{2}_\epsilon |b_{i}-a_{j}|^{2}_\epsilon}
\prod_{i=1}^{q}  
\frac{d^2 a_i d^2 b_i}{|a_i-b_i|^2_\epsilon }
\label{13}
\end{equation}%
where the function
\begin{equation}
\Phi _q\left(a,b,{\bf t}^{1},{\bf t}^{2}\right)=\prod_{i=1}^{q} \left( \frac{a_i}{b_i}\right)
^{n^{(1)}}\left( \frac{\bar{a}_i}{\bar{b}_i}\right)
^{-n^{(2)}}e^{V(a_i,t^{(1)})-V(b_i,t^{(1)})-V({\bar{a}_i},t^{(2)})+V({\bar{b}_i }%
,t^{(2)})  }  \label{Phi}
\end{equation}%
and%
\begin{equation}
V(z,t)=\sum_{m>0}t_{m}z^{m}\quad .  \label{V}
\end{equation}%
Because $\Phi _q\left(a,b,0,0\right)=1$, the tau function evaluated at ${\bf t}^{(1)}={\bf t}^{(2)}=0$ is equal
to the instanton grand partition function
$
\tau^A \left(0,0|D,\epsilon\right) = Z_{\rm inst}
$
and, for
\begin{equation}
\phi_q (a,b)=\Phi_q\left(a,b,{\bf t}^{1},{\bf t}^{2}\right)  \label{16}
\end{equation}
 we observe
\begin{equation}
 \left<  \phi \right>_{\rm inst}^A=
 \frac{\tau^A \left({\bf t}^{1},{\bf t}^{2}|D,\epsilon\right)}{\tau^A \left(0,0|D,\epsilon\right)},
 \label{17}
\end{equation}%
In the rest part of the paper we put $n^{(1)}=n^{(2)}=0$.

\paragraph{Certain correlation functions}

\paragraph{Discrete KP equations.}
If we specify the parameters as follows: 
\be\label{HMiwa}
t^{(1)}_k[\texttt{n},z] :=-\frac 1k \sum_{i=1}^N \texttt{n}_i z_i^{-k},\quad
t^{(2)}_k[\texttt{m},y]=-\frac 1k \sum_{i=1}^M  \texttt{m}_i y_i^{-k}
\ee
and denote such sets as ${\bf t}^{1}[\texttt{n},z]$ and ${\bf t}^{2}[\texttt{m},y]$,
we obtain
\be\label{Phi-omega}
\Phi_q\left(a,b,{\bf t}^{1}[\texttt{n},z],{\bf t}^{2}[\texttt{m},y]\right) =
\prod_{i=1}^N \left(\omega_q(a,b,z_i)\right)^{\texttt{n}_i}
\prod_{i=1}^M \left(\omega_q({\bar a},{\bar b},y_i)\right)^{-\texttt{m}_i}
\ee
where $\omega_q$ was defined by (\ref{3}). Let us notice that 
$\omega(a,b,z)\omega({\bar a},{\bar b},{\bar z})=|\omega(a,b,z)|^2$.

Tau function written in the variables defined by (\ref{HMiwa}) solves the so-called discrete KP equation
see \cite{JM} (and \cite{Zabrodin-2012} for the review).
If $\sigma^{(i)}(x),\,i=1.2.3$ are instanton solutions of form (\ref{3})-(\ref{stereo}), then,
for the correlation function
\[
 G_{\texttt{n}_1,\texttt{n}_2,\texttt{n}_3}(z_1,z_2,z_3):=
 \left< \left(\frac{\sigma ^{1}(z_1)+i\sigma ^{2}(z_1)}{1+\sigma ^{3}(z_1)}\right)^{\texttt{n}_1} 
 \left(\frac{\sigma ^{1}(z_2)+i\sigma ^{2}(z_2)}{1+\sigma ^{3}(z_2)}\right)^{\texttt{n}_2}
 \left(\frac{\sigma ^{1}(z_3)+i\sigma ^{2}(z_3)}{1+\sigma ^{3}(z_3)}\right)^{\texttt{n}_3}
 \right>_{\rm inst}^A
\]
one can write the discrete Hirota bilinear equation (known also as discrete KP equation):
\[
(z_2-z_3)G_{\texttt{n}_1+1,\texttt{n}_2,\texttt{n}_3}(z_1,z_2,z_3)G_{\texttt{n}_1,\texttt{n}_2+1,\texttt{n}_3+1}(z_1,z_2,z_3)
\]
\[
+(z_3-z_1)G_{\texttt{n}_1,\texttt{n}_2+1,\texttt{n}_3}(z_1,z_2,z_3)G_{\texttt{n}_1+1,\texttt{n}_2,\texttt{n}_3+1}(z_1,z_2,z_3)
\]
\be\label{Hirota-for-G}
+(z_1-z_2)G_{\texttt{n}_1,\texttt{n}_2,\texttt{n}_3+1}(z_1,z_2,z_3)G_{\texttt{n}_1+1,\texttt{n}_2+1,\texttt{n}_3}(z_1,z_2,z_3) = 0
\ee
Other sets of equations may be written for general correlation functions involving
(\ref{Phi-omega}) (this will be done in a more detailed text).


\paragraph{Densities.}

As we mentioned
the denominator in (\ref{4}) coincides with the partition
function $\Xi $\ of the neutral classical Coulomb system (CCS) in the grand
canonical ensemble with the definite temperature T (T=1 see \cite{FFS}).
\begin{equation}
\tau (0,0,0,0)=\Xi
\end{equation}%
The constant K plays the role of fugacity of the Coulomb system. The
expression (\ref{17}) also coincides with the correlation function of the
CCS (at T=1). Let us consider the instanton contribution $G^{\rm inst}\left(
x,y\right) $ in the Green function%
\begin{equation}
G\left( x,y\right) =\langle \bigtriangleup _{x}\log |\omega \left( x\right)
|,\bigtriangleup _{y}\log |\omega \left( y\right) |\rangle   \label{20}
\end{equation}%
corresponding to functional $\phi \left( \omega \right) =\bigtriangleup
_{x}\log |\omega \left( x\right) |\bigtriangleup _{y}\log |\omega \left(
y\right) |$ that is $\rho (x)\rho (y)$ with

$\rho (x)=2\pi \left( \sum_{i}\delta \left( x-a_{i}\right) -\sum_{i}\delta
\left( x-b_{i}\right) \right) $. In order to obtain this result in terms of $%
\tau $ functions we have to make Miwa transformation of times%
\begin{equation}
t_{n}^{(\alpha )}=-\frac{t}{nx^{n}}-\frac{t}{ny^{n}}  \label{21}
\end{equation}%
then we achieve
\begin{equation}
\bigtriangleup _{x}\bigtriangleup _{y}\frac{\partial }{\partial t}\left(
\prod_{i}^{q}\Phi _{0,0}(a_{i},b_{i},t^{(1)},t^{(2)})|_{t_{n}^{(\alpha )}=-%
\frac{t}{nx^{n}}-\frac{t}{ny^{n}}}\right) |_{t=0}=\bigtriangleup _{x}\log
|\omega \left( x\right) |\bigtriangleup _{y}\log |\omega \left( y\right)
|=\rho (x)\rho (y).  \label{22}
\end{equation}%
One can interpret $\rho (x)$ as the charge density. We see
\begin{equation}
G^{\rm inst}\left( x,y\right) =\langle \bigtriangleup _{x}\log |\omega \left(
x\right) |\bigtriangleup _{y}\log |\omega \left( y\right) |\rangle
_{inst}=\langle \rho (x)\rho (y)\rangle _{CCS}  \label{23}
\end{equation}%
\begin{equation}
=\frac{C\bigtriangleup _{x}\bigtriangleup _{y}\frac{\partial }{\partial t}%
\left( \tau (0,0,t^{(1)},t^{(2)})|_{t_{n}^{(\alpha )}=-\frac{t}{nx^{n}}-%
\frac{t}{ny^{n}}}\right) |_{t=0}}{\tau (0,0,0,0)}  \label{24}
\end{equation}%
Similarly to the previous way we can obtain the instanton contribution in
the more general Green function corresponding to the functional
\begin{equation}
\phi \left( \omega \right) =\bigtriangleup _{x_{1}}\log |\omega \left(
x_{1}\right) |\bigtriangleup _{x_{2}}\log |\omega \left( x_{2}\right)
|...\bigtriangleup _{x_{m}}\log |\omega \left( x_{m}\right) |  \label{25}
\end{equation}%
by
\begin{equation}
G^{\rm inst}\left( x_{1},x_{2},...x_{m}\right) =\langle \rho (x_{1})\rho
(x_{2})...\rho (x_{m})\rangle _{CCS}=  \label{26}
\end{equation}%
\begin{equation*}
=\frac{C\bigtriangleup _{x_{1}}\bigtriangleup _{x_{2}...}\bigtriangleup
_{x_{m}}\frac{\partial }{\partial t}\left( \tau
(0,0,t^{(1)},t^{(2)})|_{t_{n}^{(\alpha )}=-\frac{t}{nx_{1}^{n}}-\frac{t}{%
nx_{2}^{n}}...\frac{t}{nx_{m}^{n}}}\right) |_{t=0}}{\tau (0,0,0,0)}
\end{equation*}

\subsection{Two-component KP and the regularization (B)\label{2KP-regB}}
In this case, we have
\be\label{4B}
 \left<\phi \right>_{\rm inst}^B=\frac{
\sum_{q\geq 0}\frac{K^{q}}{(q!)^{2}}\sum_{D^{2q}}
\phi _q\left(a,b\right)
\prod_{i<j\leq q}\frac{ |a_{n_im_i}-a_{n_jm_j}|^{2}|b_{n_im_i}-b_{n_jm_j}|^{2}}
{|a_{n_im_i}-b_{n_jm_j}|^{2} |b_{n_im_i}-a_{n_jm_j}|^{2}}
\prod_{i=1}^{q}  
\frac{1}{|a_{n_im_i}-b_{n_im_i}|^2 }}
{\sum_{q\geq 0}\frac{K^{q}}{(q!)^{2}}\sum_{D^{2q}}
\prod_{i<j\leq q}\frac{ |a_{n_im_i}-a_{n_jm_j}|^{2}|b_{n_im_i}-b_{n_jm_j}|^{2}}
{|a_{n_im_i}-b_{n_jm_j}|^{2} |b_{n_im_i}-a_{n_jm_j}|^{2}}
\prod_{i=1}^{q}  
\frac{1}{|a_{n_im_i}-b_{n_im_i}|^2 }}
\ee
where $\sum_{D^{2q}}$ means $\sum_{n_1,\dots, n_q,m_1,\dots m_q \in D}$ and where $a_{nm}, b_{nm}$ are
given by (\ref{a_nmb_nm}).

One just needs to replace integrals by sums according to (\ref{UVreg-B}) in the expression (\ref{8}):
\[
\tau_{2KP} (n^{(0)},n^{(1)},n^{(2)},t^{(1)},t^{(2)}|D,h)=
\]
\begin{equation}
\langle n^{(1)},n^{(2)}|\,\Gamma \left(
t^{(1)}\right)\Gamma\left(t^{(2)}\right) \,e^{K^{\frac{1}{2}}\sum_{(k,m)\in D^2} \psi ^{(1)}(a_{km})\psi ^{\dag
(2)}({\bar a_{km}}) }\,e^{K^{\frac{1}{2}%
}\sum_{(k,m)\in D^2} \psi ^{(2)}({\bar b_{km}})\psi ^{\dag (1)}(b_{km})%
 }\,|n^{(2)}-n^{(0)},n^{(1)}+n^{(0)}\rangle   \label{8A}
\end{equation}%
where the summation range in the exponents is chosen as $0\le n,m \le L$. We obtain the enumerator in 
(\ref{4B}):
\begin{equation}
\tau^B \left({\bf t}^{1},{\bf t}^{2}|D,h\right)=
\sum_{q\geq 0}\frac{K^{q}}{(q!)^{2}}\sum_{D^{2q}}
\Phi _q\left(a,b,{\bf t}^{1},{\bf t}^{2}\right)
\prod_{i<j\leq q}\frac{ |a_{n_im_i}-a_{n_jm_j}|^{2}|b_{n_im_i}-b_{n_jm_j}|^{2}}
{|a_{n_im_i}-b_{n_jm_j}|^{2} |b_{n_im_i}-a_{n_jm_j}|^{2}}
\prod_{i=1}^{q}  
\frac{1}{|a_{n_im_i}-b_{n_im_i}|^2 }
\label{13B}
\end{equation}%
If we choose $\phi_q(a,b)=\Phi_q(a,b,{\bf t}^1,{\bf t}^2)$ we get the same relations as in the previous case A,
we replace $\left< *\right>^A_{\rm inst}$ by $\left< *\right>^B_{\rm inst}$.

\subsection{One-component KP and the regularization (B)\label{KP-regB}}
The regularization (B) can be also written as the following KP tau function
\be\label{8A'}
\tau^{\rm B}_{\rm KP}({\bf t}|D,h)=
\langle n|\Gamma(t)e^{K\sum_{D^2}\frac{\psi(a_{nm})\psi^\dag(b_{nm})}{{\bar a}_{nm}-{\bar b}_{nm}}}|n\rangle
\ee
where $a_{nm}$ and $b_{nm}$ are given by (\ref{a_nmb_nm}), and
\[
 \Gamma(t)=e^{\sum_{m>0} t_m J_m},\quad J_m=\sum_{i\in\mathbb{Z}}\psi_i \psi^\dag_{i+m}
\]
$\Gamma(t)$, fermi fields and $\Phi_q$ are the same as in subsection \ref{2KP-regA} where the second 
component is absent:
\[ 
\Phi _q\left(a,b,{\bf t}\right)=\prod_{i=1}^{q} \left( \frac{a_i}{b_i}\right)
^{n}e^{V(a_i,t)-V(b_i,t) }  ,\quad 
\Phi _q\left(a,b,{\bf t}[\texttt{n},z]\right)=\prod_i \left( \omega(z_i) \right)^{\texttt{n}_i}
\label{Phi-KP}
\]
where $V$ was defined  in (\ref{V}) and 
where ${\bf t}[\texttt{n},z]$ is the choice $n=0$ and $t_m=-\sum \texttt{n}_i z_i^m,\,m>0$.
We get the same equation (\ref{Hirota-for-G}) for
\[
G_{\texttt{n}_1,\texttt{n}_2,\texttt{n}_3}(z_1,z_2,z_3):= 
\left<\left(\omega(z_1)\right)^{\texttt{n}_1}\left(\omega(z_2)\right)^{\texttt{n}_2}
\left(\omega(z_3)\right)^{\texttt{n}_3}  \right>_{\rm inst}^B
\]
Formula (\ref{8A'}) yields the same answer as (\ref{8A}) if we put ${\bf t}^2=0$ and ${\bf t}^1={\bf t}$
(see Appendix \ref{useful-relations}).
However in the case of the one-component KP we can not construct $|\omega(z)|$ by the specialization
of the parameters $t_1,t_2,\dots$ in $\Phi$.

\section*{Acknowledgements}

The work of A.O. has been funded by RFBR grant 18-01-00273a and the RAS Program 
``Fundamental problems of nonlinear mechanics'' and 
by the Russian Academic Excellence Project~\mbox{'5-100'}.

\appendix

\section{Appendix. Useful relations\label{useful-relations}}
We use the following relations
\[
 \Gamma(t)\psi(z)=e^{V(z,t)}\psi(z)\Gamma(t),\quad
 \Gamma(t)\psi^\dag(z)=e^{-V(z,t)}\psi^\dag(z)\Gamma(t)
\]
and $\Gamma(t)|n\rangle =|n\rangle$. Then
\[
 \langle n|\psi(z_1)\psi^\dag(y_1)\cdots \psi(z_q)\psi^\dag(y_q)|n\rangle =
\prod_{i<j}^q \frac{(z_i-z_j)(y_i-y_j)}{(z_i-y_j)(y_i-z_j)}\prod_{i=1}^q 
\frac{1}{z_i-y_i}\left(\frac{z_i}{y_i}\right)^n
\]
Also
\[
 e^{\sum_{i,j}\xi_i\eta_jA_{i,j}}=
 1+\sum_{q>0}\sum_{\alpha_1>\cdots\alpha_q\atop \beta_1>\cdots >\beta_q}
 \xi_{\alpha_1}\cdots \xi_{\alpha_q}\eta_{\beta_1}\cdots 
 \eta_{\beta_q} \det\left(A_{\alpha_i,\beta_j} \right)
\]
where $\xi_i,\eta_i$ are odd variables (in our case: fermi fields with the property $\xi_i\eta_j+\eta_j\xi_i=0$
for each pair $i,j$),
and $A_{i,j}$ is (possibly infinite) matrix.

And at last
\[
 \det\left(\frac{1}{z_i-y_j}  \right)=
\prod_{i<j} \frac{(z_i-z_j)(y_i-y_j)}{(z_i-y_j)(y_i-z_j)}\prod_{i} 
\frac{1}{z_i-y_i}
\]

\section{Appendix. Bilinear identity for the$\ \protect\tau $ function}

In this section we define more general $\tau $ functions in comparison with (%
\ref{8}):
\begin{equation}
\tau (n_{1},n_{2},n,t^{(1)},t^{(2)})=\langle n_{1},n_{2}|\,\Gamma \left(
t^{(1)},t^{(2)}\right) g\,|n_{2}-n,n_{1}+n\rangle \quad ,  \label{27}
\end{equation}%
where
\begin{equation}
g=\,e^{K^{\frac{1}{2}}\int \psi ^{(1)}(a)\psi ^{\dag (2)}({\bar{a}}%
)d^{2}a}\,e^{K^{\frac{1}{2}}\int \psi ^{(2)}({\bar{b}})\psi ^{\dag
(1)}(b)d^{2}b}\quad  \label{28}
\end{equation}%
and $\Gamma \left( t^{(1)},t^{(2)}\right) $ is given by (\ref{9}).
Particularly the interesting for us $\tau $ function
\begin{equation*}
\tau (n_{1},n_{2},t^{(1)},t^{(2)})=\tau (n_{1},n_{2},0,t^{(1)},t^{(2)})
\end{equation*}

The bilinear identity is valid in the following form (see \cite{JM}). For $%
n_{1}-n_{1}^{\prime }\geq n^{\prime }-n\geq n_{2}^{\prime }-n_{2}+2$, we have%
\begin{equation}
\sum_{\alpha =1}^{2}\oint \frac{dz}{2\pi iz}\langle n_{1},n_{2}|\,\Gamma
\left( t^{(1)},t^{(2)}\right) \psi ^{(\alpha
)}(z)g\,|n_{2}-n-1,n_{1}+n\rangle  \label{29}
\end{equation}%
\begin{equation*}
\times \langle n_{1}^{\prime },n_{2}^{\prime }|\,\Gamma \left( t^{^{\prime
}(1)},t^{\prime (2)}\right) \psi ^{\dag (\alpha )}(z)g\,|n_{2}^{\prime
}-n^{\prime }+1,n_{1}^{\prime }+n^{\prime }\rangle =0
\end{equation*}%
and the integration is taken along a\ small coutour at $z=\infty $ so that $%
\oint \frac{dz}{2\pi iz}=1$.

Rewrite this ($\ref{21}$)\ we obtain
\begin{equation}
\oint \frac{dz}{2\pi iz}\left( -1\right) ^{n_{2}+n_{2}^{\prime
}}z^{n_{1}-1-n_{1}^{\prime }}e^{V(z,t^{(1)}-t^{\prime (1)})}  \label{30}
\end{equation}%
\begin{equation*}
\times \tau (n_{1}-1,n_{2},n+1,t^{(1)}-\theta \left( z^{-1}\right)
,t^{(2)})\tau (n_{1}^{\prime }+1,n_{2}^{\prime },n^{\prime }-1,t^{\prime
(1)}+\theta \left( z^{-1}\right) ,t^{\prime (2)})
\end{equation*}%
\begin{equation*}
+\oint \frac{dz}{2\pi iz}z^{n_{2}-1-n_{2}^{\prime }}e^{V(z,t^{(2)}-t^{\prime
(2)})}
\end{equation*}%
\begin{equation*}
\times \tau (n_{1},n_{2}-1,n,t^{(1)},t^{(2)}-\theta \left( z^{-1}\right)
)\tau (n_{1}^{\prime },n_{2}^{\prime }+1,n^{\prime },t^{\prime
(1)},t^{\prime (2)}+\theta \left( z^{-1}\right) )\quad ,
\end{equation*}%
where $\theta \left( z^{-1}\right) =\left( \frac{1}{z},\frac{1}{2z^{2}},...%
\frac{1}{nz^{n}},...\right) $.

An example of (\ref{22}), we have the following bilinear equations for $%
f=\tau (n_{1},n_{2},0,t^{(1)},t^{(2)})=\tau (n_{1},n_{2},t^{(1)},t^{(2)})$, $%
g=\tau (n_{1}-1,n_{2}+1,1,t^{(1)},t^{(2)})$ \ and $g^{\ast }=\tau
(n_{1}+1,n_{2}-1,-1,t^{(1)},t^{(2)})$:%
\begin{equation}
\left( D_{t_{2}^{\left( 1\right) }}-D_{t_{1}^{\left( 1\right) }}^{2}\right)
f\cdot g=0,\qquad \left( D_{t_{2}^{\left( 1\right) }}-D_{t_{1}^{\left(
1\right) }}^{2}\right) g^{\ast }\cdot f=0,  \label{31}
\end{equation}%
\begin{equation*}
\left( D_{t_{2}^{\left( 2\right) }}+D_{t_{1}^{\left( 2\right) }}^{2}\right)
f\cdot g=0,\qquad \left( D_{t_{2}^{\left( 2\right) }}+D_{t_{1}^{\left(
2\right) }}^{2}\right) g^{\ast }\cdot f=0,
\end{equation*}%
\begin{equation*}
D_{t_{1}^{\left( 1\right) }}D_{t_{1}^{\left( 2\right) }}f\cdot f-2g\cdot
g^{\ast }=0,\qquad \qquad \qquad \qquad \qquad \qquad
\end{equation*}%
where Hirota operator as usual $D_{x}\sigma \cdot \tau =\frac{\lim }{%
\varepsilon \rightarrow 0}\frac{\partial }{\partial \varepsilon }\sigma
\left( x+\varepsilon \right) \tau \left( x-\varepsilon \right) =\sigma
_{x}\tau -\sigma \tau _{x}$

\end{document}